# DESIGN TRACKING: TRACK AND REPLAY BIM-BASED DESIGN PROCESS

**Xiang-Rui Ni[1], Zhe Zheng[1], Jia-Rui Lin[1,*], Zhen-Zhong Hu[2], Xin Zhang[1]**

1  Department of Civil Engineering, Tsinghua University, Beijing, China
2  Institute for Ocean Engineering, Shenzhen International Graduate School, Tsinghua University, Shenzhen, China
* Corresponding author (lin611@tsinghua.edu.cn)

**Abstract**
Among different phases of the life cycle of a building or facility, design is of the utmost importance to ensure safety, efficiency and sustainability of the building or facility. How to control and improve design quality and efficiency has been explored for years, and more studies emerged with the popularization of Building Information Modelling (BIM). However, most of them focused on the extraction of design behaviors, while paying less attention to how a design is formed. Therefore, this study proposes an approach to tracking and replaying the BIM-based design process by integrating data logging and 4D visualization techniques. First of all, potential design behaviors and procedures are analyzed and extracted by observing how a designer designs a BIM model. Meanwhile, the required data for logging design process is defined and a relevant method to collect these data is developed based on the APIs of BIM software. Then, strategies on how to visualize different design procedures are designed and implemented via 4D visualization. Finally, a prototype system is developed based on Autodesk Revit and validated through a case study. Result shows that the proposed approach enables intuitively and interactively review of the design process, and makes it easier to understand design behaviors and even identify potential pitfalls, thus improving the design efficiency and quality.

## 1. Introduction

In the life cycle of engineering construction projects, the three core stages are the design stage, the construction stage and the operation and maintenance stage. The design stage often only takes up a small part of the time and cost of the whole project, but it plays a decisive role in the cost, quality and duration of the project [1]. Therefore, improving the efficiency and quality of the design stage is very important to ensure the success of the whole project. In order to study the specific methods to improve efficiency and quality of the design stage, it is necessary to conduct a detailed analysis of the specific process of design. The design process is of significant importance because it can reflect factors directly related to design efficiency and quality, such as the designer's ideas, behavioural habits and experience level. This information usually cannot be obtained from the design results, such as blueprints, drawings, calculation results.

Nowadays, BIM-based design has become more and more common in the architecture, engineering and construction (AEC) industry. Building Information Modelling (BIM) provides a data integration platform for the whole life cycle of buildings, recording comprehensive and diverse data. The traditional CAD-based design can only provide design results, while BIM-based design can record the detailed data of the design stage and provide valuable information for studies on the specific process of design. There have been some studies focusing on tracking the specific process of design. Yarmohammadi, etc. [2] evaluated the modelling efficiency of different designers by analysing the real-time recorded design process data, and based on this, gave a staffing plan to optimize the efficiency of the whole design team. Based on the design process information mined from the BIM log file, Pan, etc. [3] established a prediction model for design behaviour using Long Short-Term Memory neural network.

However, these studies based on data analysis had the following shortcomings. First of all, most of the studies extracted the design behaviour data from BIM software and analysed the data outside it. This process can be time-consuming, and the use of multiple data analysis software also makes the whole process very complicated. Secondly, it also took time to send feedback on the conclusion to the designers after the data analysis. The feedback including data results (data, tables, text conclusions, etc.) can be abstract so that designers cannot quickly identify and understand the specific modelling details. This makes it more difficult to match the specific design behaviour with the feedback to draw any practical conclusion [4].

To address this problem, we can use visualization technique to show the design process to the designers which provides designers with a quick way to learn and summarize experience, and the researchers with a means to observe the design process visually. In the AEC industry, especially in the studies related to construction planning, the 4D visualization technique has been used to combine the 3D data of BIM models with the schedule information to obtain the 4D analysis model of the building life cycle, so as to visually display the project progress and assist in checking the correctness and constructability of the building design [5]. There are many specific implementation schemes of 4D visualization technique [6], but most of them need to use additional software and link the BIM model with it. However, the data conversion between different kinds of software involves steps such as the unification of format and type, and the process is complicated [5].

To address the shortcomings of the previous studies, this paper designed a system for tracking and intuitively "replaying" the design process based on 4D visualization technique, which can realize the 4D visualization of the modelling process directly in the BIM software used by designers. The system can meet the requirements of quickly and accurately locating the specific design process and analysing the design behaviour characteristics. This process can transform the implicit knowledge hidden in design behaviour into explicit knowledge expressed through written experiences, providing clear suggestions and guidance for future design work [7]. This study first analysed the design behaviour of designers based on the theory of Human-Computer Interaction and summarized the data requirements of the system. This analysis not only gave corresponding data requirements, but also provided a foundation for designing specific 4D visualization strategy. Specific algorithms were then developed to collect data and realize visual effects. To prove the effectiveness of the proposed system, a prototype system based on Revit software was developed. Finally, a case study using the prototype was conducted to validate the proposed system. In the end, this paper discusses the value of the system and the deficiencies that need to be further improved.

## 2. Methodology

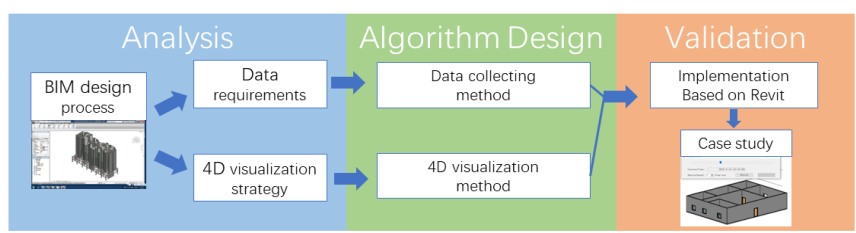

Fig. 1. Methodology of this study

As shown in Fig. 1, this study first analyses the BIM design process to summarize the data requirements of the system and give a clear goal for visualization. The specific method of 4D visualization is the core part of this system, therefore it is crucial to analyse and propose an implementation strategy of 4D visualization for specific issues of concern in this study. Specifically, this study proposed a 4D visualization strategy combining visual effects and interactive UI window. The algorithm design stage produces a data collecting method to address the data requirements and a specific method to realize 4D visualization in BIM software. Finally, this study implemented specific code based on Autodesk Revit software to verify the effectiveness of the design, and tested it with a case study in a specific BIM model.

*2.1. Analysis of BIM Design Process and Data Requirements*

The design process based on BIM is a human-computer interaction process. In this process, information flows in two directions. The designer inputs instructions through the computer input hardware, the computer then processes these instructions, makes modifications to the model, and returns the new model to the designer through the screen. The designer can then carry out the next modelling step according to the current results. Just like replaying a video, this system should enable users to see when each step was carried out on the timeline in order to meet the requirements of the basic replay function. The most important operations that make changes to the BIM model can be divided into three categories: creation, modification and deletion [8]. These are also the three types of user operations that this study focuses on and require visualization.

The core data requirement of tracking and replaying modelling process is the time series data of BIM model modifications. These time series data should contain two types of information. First of all, data that can uniquely identify model elements is required, which is usually the ID of elements in the model. The second type is the information about user operations, including the type of operation (creation, modification, deletion, etc.), the time of operation and other extra data that may be required.

Table 1. Data requirements of different types of operations

| Types of operation | Common data requirements | Different extra data requirements |
| --- | --- | --- |
| Creation | IDs, Time | -- |
| Modification |  | Specific change data. e.g., spatial position change, material change |
| Deletion |  | Data needed to reconstruct the element |

Different types of operation need different extra data (Table 1) and the complexity of the required data varies greatly. For example, the creation operation requires no extra data, while the deletion operation needs all the parameter information needed to recreate the deleted element. This is because if we want to visualize an element that has been deleted from the database, we first need to be able to reconstruct it in the model. However, the quantity and fineness of the required data are not absolute and can be adjusted according to user's demands. If it is only needed to replay the process of creation elements, or to be only accurate about the basic information of the deleted elements, such as the geometric position and size of the deleted element, it is possible to reduce the number and fineness of corresponding data requirements to simplify the data collecting process.

*2.2. Analysis of 4D Visualization Strategy*

This study only relies on BIM software for visualization, its advantage is convenience and speed, but it is also subject to mainly two aspects of limitations. Firstly, BIM software can only provide limited visual effects-related settings for visualization, and most BIM software does not allow deeper customization for graphic renderers. Secondly, additional windows need to be developed to control the visualization process and serve as data panels. Therefore, the strategy that BIM based 4D visualization should adopt is to use the visual effects provided by the software combined with interactive windows that control visualization and display relevant data.

The visual effect of "replay" can be realized through the display mode or display style of elements in BIM software. The display mode is mainly related to the current state parameter of the element, such as whether it is selected, whether it is in the hidden state, etc. BIM software uses different visual methods to highlight components in these different states, which can be used by our visualization purpose. For example, in Revit, an element is shown in blue if selected. The display style depends on the rendering options provided by the graphic renderer. For example, Revit provides display style options such as "Realistic", "Wireframe", and "Consistent Colours". These options only affect the display effect of the elements in a particular view, and do not change actual parameters of them.

The window providing interactive functions needs to control the progress of 4D visualization and dynamically adjusts the visual effects of elements in the view according to user operations. We also need windows to display some relevant information, such as the time of the current operation, the types of the elements involved and the type of operations. If the demand for data display is low, these two kinds of windows can be merged together.

*2.3. Data Collecting Method Design*

In this study, real-time data collecting method was used. Real-time data collecting can be carried out through software plugins of BIM software, which can be developed based on redeveloping method. The cost of additional software development is high and it is difficult to obtain the data in the BIM model, so the plugin based on redevelopment is more suitable. Most sorts of BIM software on the market have open Application Programming Interfaces (APIs), which usually provide rich functions for redevelopment.

The data requirements of this study include the user's operation data, and the collection of this part of data needs to be realized with the help of the *Event* system. Generally, plugins need to subscribe to events related to model modification and element change provided by BIM software to obtain relevant data when these events occur. The collected data needs to be exported and stored in a serialized way, and different data belonging to different models should be stored separately.

*2.4. 4D Visualization Method Design*

The 4D visualization strategy obtained from the previous analysis is a universal principle that can be applied to almost all kinds of BIM software. However, for each specific BIM design software, the functions and rendering parameters provided by the software are different, so the 4D visualization method in each specific software needs to be designed independently. This section provides a possible visualization scheme for Autodesk Revit.

The specific plan for visual effects of 4D visualization is shown in the table 2. Different design behaviours are presented in different ways by visual effects. A replay control window is needed to read the data files and modify the BIM model in real-time based on user actions to achieve visual effects. This window should also present relevant data.

Table 2. A possible visualization plan for Revit

| Types of operation | Creation | Modification | Deletion |
|---|---|---|---|
| Visual effect | Hide/display mode set to display | Shown in blue | Set to be transparent |
| Examples | 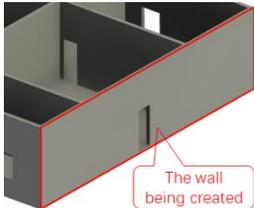 | 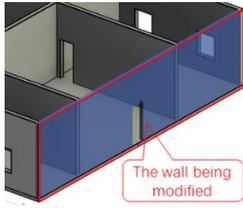 | 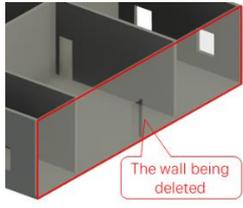 |

*2.5. Implementation and Validation*

Specific implementations and case studies are important ways to test the usability and effectiveness of a newly proposed system. This study takes Autodesk Revit software as an example to develop specific codes, collect required design process related data through the Revit plugin, and provide modelling process replay functionality. Considering the difficulty of development, the plugin developed in this study only implements replay of creation operations. The developed plugin will be used on a simple specific BIM model for specific case testing and analysis. The results of the case study are then used to evaluate the functionality, efficiency, and user experience of the plugin.

## 3. Implementation and Case Study

*3.1. Implementation*

This study developed a prototype system based on Revit to realize the basic tracking and replaying functions for BIM-based design, and verified the feasibility and effectiveness of the proposed system. The main body of the prototype system is a plugin developed for Revit 2023 on Windows.

Tracking the creation of an element requires the identification data of the element and the time of the creation operation. In Revit software, each element of an independent model file has a unique *Elementid* attribute as the identification code. This attribute can be easily obtained using APIs, and the time information can be obtained directly through the *System.DateTime* method provided by Windows. For real-time recording, the system subscribed to two events provided by Revit, *DocumentOpenedEvent* and *DocumentChangedEvent*. The former notifies the plugin after opening a model file. At this time, the plugin retrieves whether there is a data record file corresponding to the model file. If not, it creates a data recording file and prepares the file stream for subsequent writing. If there is a matching data record file, it opens the file and prepare the file stream. The latter event will notify the plugin after the model changes, and will provide a list of *ElementId* of all newly added elements. The plugin then integrates these data together with the time information into a structured string and stores it in the data file.

The data recording file of the prototype system adopts the txt format, mainly considering that the data to be recorded during the modification and deletion operations is relatively complex. This may cause the length and complexity of the records to vary greatly, so a certain degree of structure is sacrificed to ensure flexibility and versatility which may benefit further development.

The "replay" visual effect is achieved by changing the hide/display mode of elements. For a specific point in time, elements that have been created are displayed, and elements that have not yet been created are hidden. Over time, hidden elements change to the displayed state one by one in the time order of creation. In Revit, the hide and display of elements are set by the viewport (*View*). The hide/display of elements are designated by a specific *View*. In other word, if an element is hidden in one *View*, it is still displayed in other *Views*.

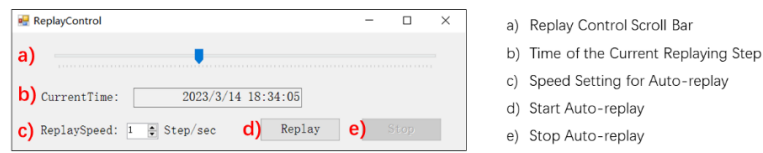

Fig. 2. Replay control interface

To provide users with more control options over the replay function, the plugin includes a dialog interface (Fig. 2). Clicking the corresponding button on the ribbon menu opens the replay control interface and all elements are set to be hidden to take the user back to the starting point of modelling. All playable steps are distributed evenly on the scroll bar (a) in time order. The scroll bar is both a timeline and a "video progress bar". Users can drag the handle at will and observe the changes of the model. The time of the handle position is displayed in the textbox (b). Users can also let the plugin play automatically by setting the playback speed (c) and press *Replay* (d). It will play the set number of steps every second. Automatic replay can be stopped by pressing *Stop* (e) at any time, and it will also stop automatically at the end of the replay progress. When you close the interface, the hide/display state of all elements in the view will remain. If you need to restore all the elements of the model to the display state, you need to drag the scroll bar to the end and then close the interface.

*3.2. Case Study*

This study takes a simple BIM model as an example to verify the effectiveness of the prototype system. The model has two floors, including common components such as walls, slabs, roof, doors, windows, stairs. Figure 3 shows three moments in the actual replay process and the complete model.

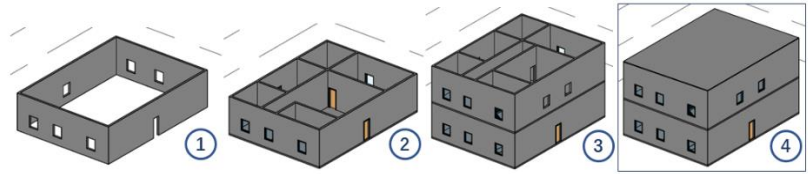

Fig. 3. Three moments in the replay process and the complete model

The test proves that the plugin can realize the playback function well and give users sufficient control over the replay function. During the replay, designers can review their operations and summarize experience to improve design efficiency. For example, most elements on the second floor of the case model can be created quickly by copying elements from the ground floor. If designers discovered that they had been creating elements one by one, they can summarize experience and adopt more efficient methods in future modelling tasks.

## 4. Conclusion

Current BIM-based design process data analysis is not intuitive and the existing visualization scheme is complex. This study proposes a tracking and playback system for design behaviour to address these problems. The system collects the user's design process data in real time and intuitively displays the design process in BIM software. The system provides interactive control over the replay process and basic information about the current replay progress. The effectiveness of the system is verified by the development of a prototype system based on Revit and a case study on a simple BIM model. Designers can review their design process step by step and discover which specific operations can be optimized for better efficiency and equality. The recorded data file can also provide basic data for future research, and the plugin can also be improved to record more types of data such as mouse movement, keyboard input and executed commands.

The system still has two deficiencies. Firstly, the real-time data collection method poses a problem. For models that have been partially completed elsewhere, the plugin can only replay the modelling process after using the plugin. In future research, data sources can be supplemented through log files and other means. Another disadvantage is that the current system only focuses on the creation, editing and deletion of elements, while there are many other design behaviours worthy of attention, such as perspective change, command input, etc. More diverse and detailed information of user operations should be included in future studies.

**Acknowledgements**

This work was supported by the University-Industry Collaborative Education Program (Grant No. 202102113020).